\documentclass[10pt,prl,aps,twocolumn,showpacs,superscriptaddress]{revtex4}
\usepackage{graphicx}
\usepackage{amssymb}
\usepackage{bm}% bold math

\begin{document}

\title{Impurity induced spin texture in quantum critical 2D
antiferromagnets}

\author{Kaj H. H\"oglund}
\affiliation{Department of Physics, {\AA}bo Akademi University,
Porthansgatan 3, FI-20500, Turku, Finland}

\author{Anders W. Sandvik}
\affiliation{Department of Physics, Boston University, 590
Commonwealth Avenue, Boston, Massachusetts 02215}

\author{Subir Sachdev}
\affiliation{Department of Physics, Harvard University, Cambridge,
Massachusetts 02138}

\date{\today}


\begin{abstract}
We describe the uniform and staggered magnetization distributions
around a vacancy in a quantum critical two-dimensional $S=1/2$  
antiferromagnet.
The  distributions are delocalized across the entire sample with a  
universal
functional form  arising from an uncompensated Berry phase. The  
numerical results,
obtained using quantum Monte Carlo simulations of the Heisenberg model  
on bilayer
lattices with up to $\approx 10^5$ spins, are in good agreement with  
the proposed
scaling structure. We determine the exponent $\eta'=0.40 \pm 0.02$,  
which governs
both the staggered and uniform magnetic structure away from the  
impurity and
also controls the impurity spin dynamics.
\end{abstract}

\maketitle

Some of the most interesting physics of strongly interacting quantum
systems arises in their response to impurities. Metallic systems
exhibit the Kondo effect, and much rich physics has been discovered
in their response to a magnetic impurity which carries a localized
spin $S$. In contrast, Mott insulators have a particularly rich
response to {\em non-magnetic\/} impurities. For example, the
spin-gap compound CuGeO$_3$, which consists of dimerized pairs of
$S=1/2$ Cu ions locked into $S=0$ valence bonds, acquires magnetic
order upon replacing a very small density of the Cu with spinless Zn
ions \cite{cuge}; it is believed that an unpaired Cu spin is
localized in the vicinity of the Zn impurity, and behaves like a
localized $S=1/2$ moment. The cuprate superconductors have also seen
a variety of studies \cite{bobroff,ouazi,yazdani,pan} of the spin
and charge correlations in the vicinity of Zn ions replacing the Cu
ions within a superconducting layer; here there is also an unpaired
spin, but its spatial distribution and dynamics are not fully
understood.

This paper will explore the effect of a non-magnetic impurity on a
quantum critical Mott insulator, at the boundary between a
magnetically ordered and a spin-gap state. We will describe the fate
of the spatial magnetic structure of the localized impurity in the
spin-gap state, as this state is tuned to the quantum critical
point. We find that the strongly-interacting gapless excitations in
the bulk lead to a non-trivial and universal spatial form, with
power-law decay of spin correlations away from the impurity (analogous
to skyrmions induced by dopants in the N\'eel state \cite{marino}).
We will present new numerical and analytic results on the spatial spin
distribution, building upon the scaling structure proposed in an
earlier field-theoretical analysis \cite{science,qimp2}. Our results
are of relevance to Mott insulators which can be tuned across the
quantum phase transition. They also shed light on the cuprates,
which are are in the vicinity of a magnetic ordering quantum phase
transition.

\begin{figure}
\includegraphics[width=7.0cm,clip]{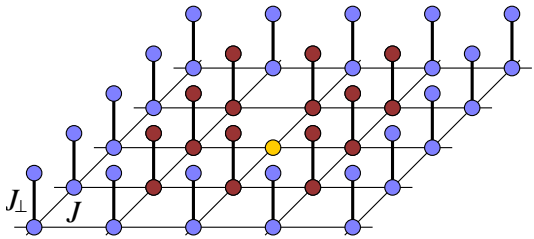}
\caption{(Color online) An $L=5$ incomplete bilayer model with a
vacancy. The single unpaired central spin (yellow) constitutes frame
$R=0$. Frames $R=1$ and $R_{\text{max}}=2$ consist, respectively, of
the surrounding red and blue sites.} \label{fig:fig1}
\end{figure}

In our numerical investigations of a model spin-gap Mott insulator,
we consider the spin-$1/2$ Heisenberg antiferromagnet on a bilayer
lattice. It is defined by the Hamiltonian
\begin{equation}
H=J\sum _{\left <i,j\right >}\mathbf{S}_{1,i}\cdot
\mathbf{S}_{1,j}+J_{\perp}\sum _{i}\mathbf{S}_{1,i}\cdot
\mathbf{S}_{2,i},
\end{equation}
where $\mathbf{S}_{a,i}$ is a spin-$1/2$ operator at site $i$ on
layer $a=1,2$, and $\left <i,j\right >$ denotes a pair of
nearest-neighbor sites on an $L\times L$ open-boundary square
lattice. With intralayer interactions in only one of the layers (a
Kondo lattice), as shown in Fig.~\ref{fig:fig1}, the model has a
quantum critical point when the ratio $g=J_{\perp}/J$ is
$g_c=1.3888(1)$ \cite{wang}, with the spin-gap state present for $g>
g_c$, and an antiferromagnetically ordered state for $g<g_c$. There
is convincing evidence \cite{matsumoto,wang} that this quantum
critical point is described by the Wilson-Fisher fixed point of
$\varphi^4$ field theory with O(3) symmetry in 3 spacetime
dimensions. Here, our discussion is conveniently presented in terms
of the fixed-length formulation of this field theory, which is the
O(3) non-linear sigma model of the field unit length field ${\bf
n}(r, \tau)$ (${\bf n}^2 = 1$), representing the local orientation
of the antiferromagnetic order, with action
\begin{equation}
\mathcal{S}_{\rm bulk} = \frac{1}{2 \widetilde{g}} \int d\tau \int
d^2 r \left[ (\partial_\tau {\bf n} )^2 + c^2 ( \nabla_r {\bf n} )^2
\right],
\end{equation}
where $\tau$ is imaginary time, $r$ is the spatial co-ordinate, $c$
is the spin-wave velocity, and the coupling $\widetilde{g}$ is a
monotonic function of $g$. At the quantum critical point at
$\widetilde{g} = \widetilde{g}_c$, the correlations of the
antiferromagnetic order are characterized by the power-law decay
\begin{equation}
\langle {\bf n} (r, \tau) \cdot {\bf n} (0,0) \rangle \sim (r^2 +
c^2 \tau^2)^{-(1+\eta)/2}, \label{eta}
\end{equation}
with the exponent $\eta \approx 0.04$ \cite{compostrini} of the
Wilson-Fisher fixed point.

Let us now add a non-magnetic impurity to the above systems. For the
lattice Hamiltonian $H$, we create a vacancy by removing a single
spin. In a finite size $L \times L$ system, this leads to a twofold
degenerate $S= 1/2$ ground state. We take $L$ odd and place the
vacancy at the center of the top layer, as shown in
Fig.~\ref{fig:fig1}. We choose the ground states to be eigenstates
of the total $S^z$; in either of these states, all the $\langle
{S^z}_{1,2,i} \rangle$ are non-zero even in zero applied field
(which we assume throughout), and exhibit an interesting spatial
structure that we will describe.

Next, we add the impurity to the field theory. The unpaired spin
near the origin leads to a net uncompensated Berry phase between the
antiferromagnetically oriented spins, adding an impurity term to the
action \cite{qimp2}
\begin{equation}
\mathcal{S}_{\rm imp} = i S \int d\tau {\bf A} [ {\bf n} (0,\tau)]
\cdot \frac{d {\bf n} (0, \tau)}{d \tau},
\end{equation}
which depends only upon the orientation of the order parameter at
$r=0$. Here $S=1/2$ is the unpaired spin associated with the
non-magnetic impurity, and ${\bf A}$ is the Dirac monopole function
in spin space with $\nabla_{{\bf n}} \times {\bf A} = {\bf n}$. It
was argued in Refs.~\cite{science,qimp2} that $\mathcal{S}_{\rm
bulk} + \mathcal{S}_{\rm imp}$ universally describes the quantum
impurity near $g=g_c$. In the infinite system, the critical spin
correlations are characterized by a different ``boundary" exponent at
$r=0$:
\begin{equation}
\langle {\bf n} (0, \tau) \cdot {\bf n} (0,0) \rangle \sim
|\tau|^{-\eta'}. \label{etap}
\end{equation}

A crucial property of the field theory $\mathcal{S}_{\rm bulk}+
\mathcal{S}_{\rm imp}$ is that, like the lattice model, for $g \geq
g_c$ the ground state has total spin $S$. This means that for the
conserved Noether magnetization density $Q_z (r)$, associated with
the symmetry of O(3) rotations of the action, has a non-vanishing
expectation value even in zero field, and obeys
\begin{equation}
\left\langle \int d^2 r Q_z (r) \right\rangle = S \label{int}
\end{equation}
in the ground state with maximum spin projection in the $z$
direction. While this spin is localized in the spin-gap state with
$g>g_c$, there is a transition to a delocalized critical state at
$g=g_c$ in which (as we describe below) the spatial extent of the
magnetization is set only by the system size. Consequently, in an
infinite system at $g=g_c$ we have $\langle Q_z (r) \rangle = 0$ at
all $r$ even though Eq.~(\ref{int}) is obeyed \cite{curiefoot}.

A proper analysis of the delocalized spin texture in the ground
state at $g=g_c$ requires imposition of a finite size $L$. We found
from our numerical results, described below, that the spin
distribution quite accurately obeys the scaling form
\begin{equation}
\left\langle Q_z (r) \right\rangle = L^{-2} \Phi_{Q} ( r/L),
\label{phiq}
\end{equation}
where $\Phi_{Q} (y)$ is a universal function (with no arbitrary
scale factors) obeying $\int d^2 y \Phi_{Q} (y) = S$. The same
scaling form also emerges from a renormalization group analysis of
the field theory $\mathcal{S}_{\rm bulk} + \mathcal{S}_{\rm imp}$,
along with explicit results for $\Phi_{Q} (y)$, and these will be
described elsewhere. More useful here is the behavior of the scaling
function as $y \rightarrow 0$, which describes the spin distribution
in the vicinity of the impurity. A key feature of the theory
\cite{science} is that {\it both} the uniform and staggered spin
operators of the bulk theory transmute into the {\em same}
``boundary" operator as they approach the impurity. Here we are
using ${\bf n} (0,\tau)$ to represent this boundary operator, and so
the scaling dimensions implicit in Eqs.~(\ref{etap}, \ref{phiq})
suggest the operator product expansion $\lim_{r \rightarrow 0} {\bf
Q} (r, \tau) \sim |r|^{-2+\eta'/2} {\bf n} (0, \tau)$, which in turn
implies that
\begin{equation}
\Phi_Q (y \rightarrow 0) \sim y^{-2+\eta'/2}. \label{y1}
\end{equation}
Similarly, we can also examine the distribution of staggered spin
density, which is encoded in the spatial distribution of the order
parameter ${\bf n}$. From Eq.~(\ref{eta}) we deduce the scaling form
\begin{equation}
\left\langle n_z (r) \right\rangle = L^{-(1+\eta)/2} \Phi_{n} (
r/L). \label{phin}
\end{equation}
Now the operator product expansion to the same boundary operator is
$\lim_{r \rightarrow 0} {\bf n} (r, \tau) \sim
|r|^{-(1+\eta-\eta')/2} {\bf n} (0, \tau)$, and this leads to
\begin{equation}
\Phi_n (y \rightarrow 0) \sim y^{-(1+\eta-\eta')/2}. \label{y2}
\end{equation}
Unlike $\Phi_Q$, the integral of $\Phi_n$ is not quantized, and its
overall scale is non-universal. Note that the theoretical results in
Eqs.~(\ref{phiq},\ref{y1},\ref{phin},\ref{y2}) are tightly
constrained, dependent only upon a single exponent $\eta'$ in
addition to the standard bulk critical correlation function exponent
$\eta \approx 0.04$. The value of $\eta'$ has previously been
estimated in the time domain, using Eq.~(\ref{etap}) \cite{troyer}.
We will show below that the numerics confirm that the spatial
structure is also governed by this exponent, which we evaluate to
higher precision than previously.

In order to numerically study how the impurity-induced total
magnetization $S^z=\sum_i S^z_i = \pm 1/2$ is distributed in the
system at $T=0$, the lattice is decomposed into ``frames''
surrounding the vacancy, as illustrated in Fig.~\ref{fig:fig1}. For
each frame $R$ we determine the uniform and staggered
magnetizations, respectively, defined by,
\begin{eqnarray}
M_0(R)&=&\left <s \sum _{i\in R}(S^z_{1,i}+S^z_{2,i})\right
> ,\label{modef}\\
M_\pi (R)&=&\left <s\sum _{i\in
R}(-1)^{x_i+y_i}(S^z_{1,i}-S^z_{2,i})\right >, \label{mpidef}
\end{eqnarray}
where $s=2S^z=\pm 1$ is included in order to make the contributions
positive for both $S^z=\pm 1/2$ states. Expectation values are
calculated in $T>0$ quantum Monte Carlo simulations utilizing the
stochastic series expansion algorithm~\cite{sse}. Ground state
results are obtained by choosing a sufficiently low temperature for
each $L$. For the largest lattice we have studeied, $L=257$,
temperatures as low as $T/J\approx 10^{-3}$ are required for
satisfactory $T\to 0$ convergence \cite{csc}.

\begin{figure}
\includegraphics[width=7.0cm, clip]{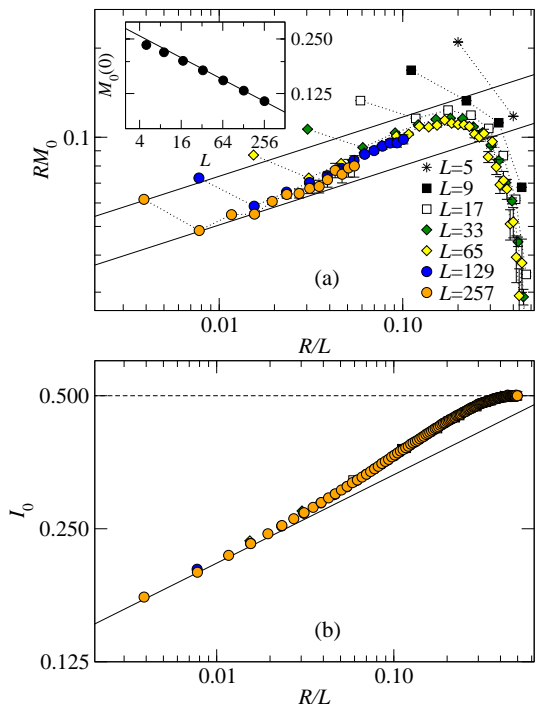}
\caption{The uniform frame magnetization $M_0(R)$ (a) and the
corresponding integrated quantity $I_0(R)$ (b) plotted according to
the predicted scaling laws for several system sizes. In the inset of
(a) we show the magnetization at $R=0$, from which we obtain the
exponent $\eta'=0.40\pm 0.02$. All other lines are shown using this
value for the exponent. In (b) the data collapse is so tight that the
$L=257$ points almost completely hide the data for smaller $L$.}
\label{fig:fig2}
\end{figure}

The uniform frame magnetization is related to the magnetization
density $Q_z$ by $M_0(R) \propto R Q_z (R)$ for $R>0$. Hence,
according to the scaling forms (\ref{phiq},\ref{y1}), we should have
for $R/L$ small
\begin{eqnarray}
M_0(R) & \sim & \frac{1}{R} \left (\frac{R}{L}\right )^{\eta^\prime
/2},\label{mag}\\
I_0(R) & \sim & \left ({\frac{R}{L}}\right
)^{\eta^\prime /2}, \label{magsum}
\end{eqnarray}
where $I_0(R)$ is the integrated frame magnetization,
\begin{equation}
I_0 (R) = \sum_{r=0}^R M_0(r), \label{i0isum}
\end{equation}
which has to be exactly $1/2$ at the edge of the lattice, i.e., for
$R_{\text{max}}=(L-1)/2$. The uniform magnetization results are
shown versus $R/L$ in log-log plots in Fig.~\ref{fig:fig2}. The
upper panel shows the frame magnetization, which is seen to collapse
onto a single curve for $L \agt 17$, with the exception of the $R=1$
points which scale with a different prefactor. Power-law behavior is
seen for small $R/L$, including also the $R=1$ data which approach
such behavior for larger $L$. For a fixed $R$, $I_0(R)$ decays
according to Eq.~(\ref{magsum}) as $L^{-\eta'/2}$. This should hold
also for $R=0$, because $I_0(R_{\rm max})=1/2$.
The inset of the upper panel of Fig.~\ref{fig:fig2}
shows the scaling of the $R=0$ magnetization, which in fact gives us
the most precise determination of the exponent; $\eta' = 0.40 \pm
0.02$. We use this value of $\eta'$ to draw the lines shown in the
main panels of Fig.~\ref{fig:fig2}. The integrated magnetization
$I_0(R)$, shown in the lower panel, scales even better than $M_0$,
with also data for the smallest lattices falling on the same curve.

The theory is expected to capture accurately the behavior for large
lattices, far from the impurity and the edges, i.e., for large $R$
but small $R/L$. It is therefore quite remarkable that even  the
$M_0(R=2)$
data fall on the common scaling curve and that the integrated
magnetization
scales almost perfectly even for very small lattices. The asymptotic
power-law
behavior is closely approached below $R/L \approx 0.02$.

\begin{figure}
\includegraphics[width=7.0cm, clip]{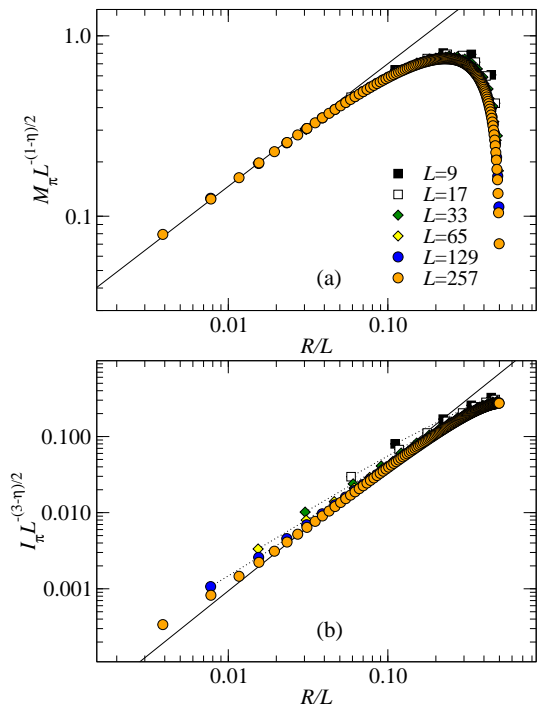}
\caption{Size and distance scaling of the staggered frame
magnetization $M_\pi(R)$ (a) and the corresponding integrated
quantity $I_\pi(R)$ (b). The over-all $L$ dependence,
$L^{-(1-\eta)/2}$ and $L^{-(3-\eta)/2}$ for $M_\pi$ and $I_\pi$,
respectively, has been divided out. The lines in (a) and (b) have
slopes $0.68$ and $1.68$, respectively, corresponding to the value
of the exponent $\eta'=0.40$ extracted from the $M_0(0)$ results in
Fig.~\ref{fig:fig2}.} \label{fig:fig3}
\end{figure}

Next we consider the staggered component (\ref{mpidef}) of the
impurity induced texture and the corresponding integral
\begin{equation}
I_\pi (R) = \sum_{r=0}^R M_\pi (r). \label{ipisum}
\end{equation}
Eqs.~(\ref{phin},\ref{y2}) imply for small $R/L$
\begin{eqnarray}
M_\pi (R)L^{-(1-\eta)/2}
   & \sim & \left (\frac{R}{L}\right)^{(1+\eta'-\eta)/2} ,  \\
I_\pi (R)L^{-(3-\eta)/2} & \sim & \left
(\frac{R}{L}\right)^{(3+\eta'-\eta)/2} .
\end{eqnarray}
Results are shown in Figs.~\ref{fig:fig3}. We observe good data
collapse of $M_\pi$ for all $L$, and the slope vs $R/L$ agrees very
well with the value $\eta'=0.40$ determined above from the uniform
magnetization. In this case the integrated quantity shows
substantial subleading size corrections for small $R$, but an $L \to
\infty$ approach to the power-law behavior shown by the line appears
very plausible.

In conclusion, we have presented analytical and numerical results
for the uniform and staggered components of the spin texture induced
by a static vacancy in a 2D quantum antiferromagnet at its quantum
critical point. The theory predicts scaling functions with
asymptotic power-law behaviors, which are very well reproduced by
the numerics. We have determined the value $\eta'=0.40 \pm 0.02$ for
the single exponent governing the asymptotic behavior of both the
uniform and staggered structure. This exponent characterizes the
influence of the Berry phase in $\mathcal{S}_{\rm imp}$ and so,
unlike the bulk theory, cannot be related to the exponent of any
classical theory in one higher dimension. Our results support a
central property of the ``boundary" critical field theory
\cite{science}: the bulk operators for the staggered and uniform
magnetizations transmute into the same boundary spin operator as
they approach the impurity. Earlier time domain studies
\cite{troyer} yielded $\eta'=0.37 \pm 0.05$ \cite{troyer}; the good
agreement between the two approaches provides strong evidence that a
single exponent indeed governs both the temporal and spatial impurity
effects \cite{science}. In addition to extracting the asymptotic  
power-law,
our numerical calculations also give the full scaling functions for  
arbitrary
distance from the impurity. We note that the integrated effects of the  
impurity
are much stronger at criticality than in the symmetry-broken N\'eel  
state, where
the induced disturbance of the magnetic structure around the impurity  
decays
asymptotically as $r^{-3}$ \cite{luscher}.

NMR \cite{bobroff,ouazi} and STM \cite{yazdani,pan} experiments have
probed the magnetization distributions around an impurity. Although
these systems are not at a quantum critical point, we hope similar
experiments in related system will be in the quantum critical
regime, allowing measurements of the exponent $\eta'$ using some of
the observables discussed in Ref.~\onlinecite{science}.

This work was supported by the NSF under grants No.~DMR-0513930
(AWS) and ~DMR-0537077 (SS). A travel grant awarded by the Finnish
Academy of Science and Letters from the Vilho, Yrj\"{o}, and Kalle
V\"{a}is\"{a}l\"{a} Foundation is also gratefully acknowledged
(KHH). Part of the simulations were performed at the CSC, the Finnish
IT Center for Science.

\null


\begin{thebibliography}{00}
\bibitem{csc}
The memory and CPU time needed for these computations scale as $N/T$.
At the lowest temperature, $T/J \approx 0.00117$, considered for the 
largest  lattice, $L=257$, approximately 3GB of RAM was required. It 
is currently  difficult to go to lower $T$ and/or larger $L$.
\end{thebibliography}
\end{document}